\documentclass[aps,prl,reprint,showpacs,superscriptaddress]{revtex4-1}

\usepackage{amsmath}
\usepackage{epsfig}
\usepackage{epstopdf}
\usepackage{textcomp}
\usepackage{color}

\begin{document}

\title{Exciton mapping at subwavelength scales in two-dimensional materials}

\author{Luiz H. G. Tizei}
\affiliation{Nanotube Research Center, National Institute of Advanced Industrial Science and Technology (AIST), Tsukuba 305-8565, Japan}

\author{Yung-Chang Lin}
\affiliation{Nanotube Research Center, National Institute of Advanced Industrial Science and Technology (AIST), Tsukuba 305-8565, Japan}

\author{Masaki Mukai}
\affiliation{JEOL Ltd., 3-1-2 Musashino, Akishima, Tokyo 196-8558, Japan}

\author{Hidetaka Sawada}
\affiliation{JEOL Ltd., 3-1-2 Musashino, Akishima, Tokyo 196-8558, Japan}

\author{Ang-Yu Lu}
\affiliation{Physical Science and Engineering Division, King Abdullah University of Science and Technology, Thuwal, 23955-6900, Kingdom of Saudi Arabia}

\author{Lain-Jong Li}
\affiliation{Physical Science and Engineering Division, King Abdullah University of Science and Technology, Thuwal, 23955-6900, Kingdom of Saudi Arabia}

\author{Koji Kimoto}
\affiliation{National Institute for Materials Science (NIMS), 1-1, Namiki, Tsukuba, Ibaraki, 305-0044, Japan}

\author{Kazu Suenaga}
\email{suenaga-kazu@aist.go.jp}
\affiliation{Nanotube Research Center, National Institute of Advanced Industrial Science and Technology (AIST), Tsukuba 305-8565, Japan}

\date{\today}

\begin{abstract}

Spatially resolved EELS has been performed at diffuse interfaces between MoS$_2$ and MoSe$_2$ single layers. With a monochromated electron source (20 meV) we have successfully probed excitons near the interface by obtaining the low loss spectra at the nanometer scale. The exciton maps clearly show variations even with a 10 nm separation between measurements; consequently the optical bandgap can be measured with nanometer-scale resolution, which is 50 times smaller than the wavelength of the emitted photons. By performing core-loss EELS at the same regions, we observe that variations in the excitonic signature follow the chemical composition. The exciton peaks are observed to be broader at interfaces and heterogeneous regions, possibly due to interface roughness and alloying effects. Moreover, we do not observe shifts of the exciton peak across the interface, possibly because the interface width is not much larger than the exciton Bohr radius.

\begin{description}
\item[PACS numbers] 79.20.Uv, 
 71.35.-y,  
71.35.Cc,  
68.37.Ma 
\end{description}

\end{abstract}

\maketitle

\textbf{Note:} this manuscript has been published Phys. Rev. Lett: L. H. G. Tizei, \textit{et al}, \textit{Phys. Rev. Lett.}, \textbf{114}, 107601 (2015). This is the last submitted version, before proof editions.

The extraordinary and unexpected properties of two-dimensional materials make them a rich ground for the observation of new physical phenomena and, also, a probable component in future devices. Transition metal dichalcogenides (TMD), such as MoS$_2$ or WSe$_2$, have been shown to suffer a transition from an indirect-gap semiconductor in bulk to a direct semiconductor in single layers \cite{Mak2010, Splendiani2010}. As in graphene, the two distinct valleys in momentum space can be used to carry information in TMD. Moreover, spin-orbit coupling splits the valence band of these materials, opening the possibility to another parameter to control carriers \cite {Xiao2012}. The valence band splitting results in two excitonic states, which have been observed in photoelectron spectroscopy \cite{Coehoorn1987} and optical absorption \cite{Molina-Sanchez2013}. These special properties of two-dimensional TMD make them interesting candidates for applications in optoeletronics, particularly at the nanoscale. To understand, characterize and improve these materials it is fundamental to observe their excitations at their typical length scales. For excitons, the scale (its Bohr radius) may vary greatly, from 2.8 nm in GaN \cite{Ramvall1998} to over 1 $\mu$m in highly-excited Rydberg excitons in Cu$_2$O \cite{Kazimierczuk2014}. In MoS$_2$ and MoSe$_2$ theory predicts the exciton wavefunction extent to be of the order of 7 nm \cite{Molina-Sanchez2013} and 3 nm \cite{Ugeda2014}, respectively. Therefore, much can be learned by probing them at the nanometer scale, far below the wavelength of light emitted by these excitons (of the order of 600 nm). The optical bandgaps are expected to be around 1.6 and 1.9 eV for MoSe$_2$ and MoS$_2$, respectively. 

Standard optical techniques, such as photoluminescence, cannot probe excitons at the subwavelength limit due to the diffraction limit. Traditional electron energy loss spectroscopy in transmission mode do allow mapping at the required scale but are limited to excitations above a few electron-volt due to the large zero loss peak tail \cite{Rafferty1998, Egerton2011, Arenal2005, Gloter2003, Nelayah2007}. High resolution EELS allows measurements of vibrational modes of molecules (better than 10 meV resolution), but without any spatial resolution \cite{Ibach1982, Fujikawa1995, Terauchi1997}. Electron monochromators fitted to electron microscopes have allowed spatially-resolved measurements with a spectral resolution of the order of 100 meV \cite{Gianluigi2013, Bosman2014}. Although this gives access to EELS experiments in the visible range, the zero-loss tail still hinders measurements. Recently, a new generation of electron monochromators have allowed measurements of phonons in hBN with some spatial resolution \cite{Krivanek2014}. 

\begin{figure}
\includegraphics[width=0.6\columnwidth]{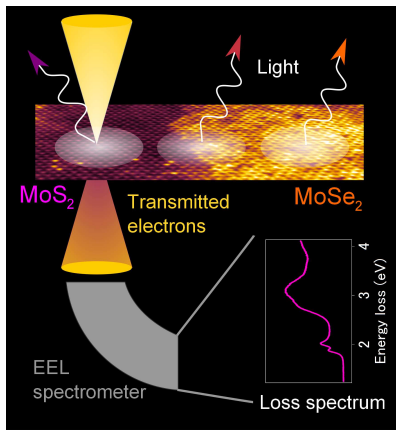}
\caption{\label{figure1} EELS experiments have been performed in scanning transmission mode where a narrow electron beam (yellow cone, width $R \sim 1$ nm) is used to probe excitons across a diffuse interface.  For the pure regions, the area probed will be limited by the size of the exciton wavefunction (represented by ellipses ), as it is larger than $R$, and delocalization effects. Interestingly, using a narrow electron probe allows access to the energy-loss spectrum of materials at scales far below the wavelength of the emitted light (arrows in the drawing).}

\end{figure}

\begin{figure}
\includegraphics[width=1\columnwidth]{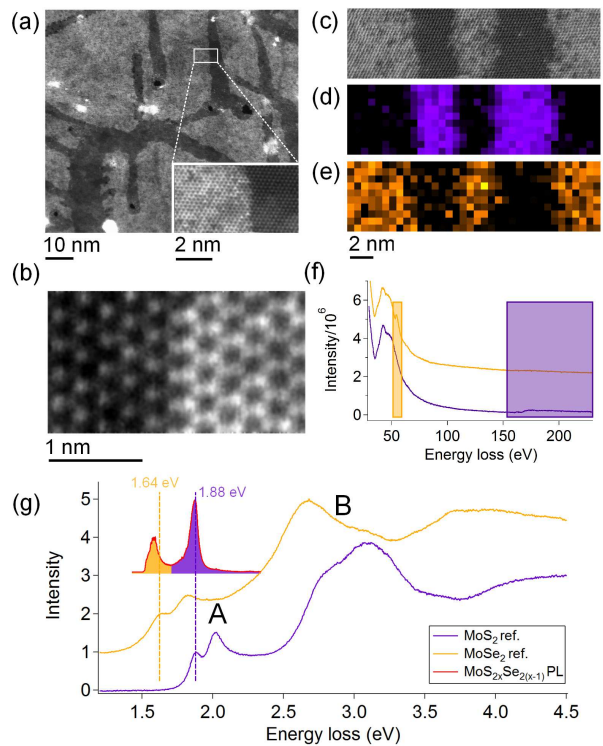}
\caption{\label{figure2} (a) HAADF images of a MoS$_{2(1-x)}$Se$_{2x}$  layer. (b) Example of a region with a sharper interface, which can be found locally. (c-f) Chemical mapping of a nanometrically mixed region of MoS$_2$ and MoSe$_2$.(c-e) shows the HAADF,  the S $L$ map, the Se $M$ map and the S $L$ and Se $M$ edges respectively. (e) Low loss EEL spectra of MoS$_2$ (purple) and MoSe$_2$ (orange). Two main features are seen. The first one, marked A, is associated with excitons. It is split in two energy-loss peaks due to spin-orbit coupling \cite{Molina-Sanchez2013}. The second feature, marked B, is also associated with an exciton. The red curve is a photoluminescence spectrum taken from the heterogeneous sample shown in (a)}
\end{figure}

In this letter we demonstrate that mapping excitons in subwavelength scales is possible (Fig. \ref{figure1}). As an example, we have measured the spatial variation of excitons in MoS$_2$/MoSe$_2$ interfaces in single two-dimensional heterogeneous layers of MoS$_{2(1-x)}$Se$_{2x}$ ($0\leq x\leq 1$) . We observed that exciton peaks are broader at interfaces, possibly due to interface roughness. This was made possible by performing electron energy loss spectroscopy (EELS) experiments using a monochromated 30 keV electron beam with a full width at half maximum (FWHM) of the order of 20 meV in a scanning transmission electron microscope with the sample at 150 K. The size of the electron beam was 1 nm. In our experiments we have observed freestanding MoS$_{2(1-x)}$Se$_{2x}$ single layers of TMD which allow high spatial resolution. These layers are heterogenenous, with regions rich in S or Se, as shown in the high angle annular dark (HAADF) image in Fig. \ref{figure2} (a). Locally, regions with sharp interfaces can be found in the nanometer scale (\ref{figure2}(b)). However, more generally, the interfaces between different regions in these layers are not abrupt, as show in Fig. \ref{figure2}(c-f). Still, regions consisting mainly of MoS$_2$ and MoSe$_2$ can be located.

In Fig. \ref{figure2}e the EEL spectra of pure single-layer MoS$_2$ and MoSe$_2$ measured with a monochromated electron beam are shown. Losses below 2 eV are discernible without deconvolution. Two sharp peaks between 1.5 and 2.1 eV (A exciton) and a broader peak at around 3.0 eV (B exciton) exciton are observed. The two A peaks are perceptible for both materials. The first A exciton and the splitting to the second one are 1.88 eV (1.64 eV) and 0.14 eV (0.19 eV) for MoS$_2$ (MoSe$_2$). The MoS$_2$ and MoSe$_2$ excitons measured using photoluminescence experiments appear at 1.93 eV \cite{Hong2014} and 1.63 eV \cite{Mann2014} at 77 K. Our measurements for the energy position of the excitons agree with those measured in direct-gap single layers using optical techniques. Unlike optical experiments, the electron source can excite higher losses. For this reason the B exciton at around 3 eV (2.7 eV) for MoS$_2$ (MoSe$_2$) can also be observed, which was theoretically predicted \cite{Molina-Sanchez2013}. For a thin anisotropic system, such as a two-dimensional monolayer, an EEL spectrum is directly related to the parallel (relative to the surface of the monolayer) dielectric function, $\epsilon_\parallel(\omega)$ \cite{Kociak2001, Taverna2002}. For this reason, in such conditions, low loss EELS measures the optical absorption of the system \cite{Kociak2001}.

The prominent interest in EELS based on scanning electron microscopy is the ability to detect spatial variations of the excitations. In a typical EELS experiment (Fig. \ref{figure1}), the electron beam is scanned over the sample and a spectrum of the energy lost by electrons provides information about various excitations and, in the low loss range, reflects the local density of optical states \cite{Abajo2008, Abajo2010}. If the size of the electron beam ($R$) is sufficiently small, the highest possible spatial resolution will be limited by the size of the excitation created and delocalization effects. Also, subwavelength spatial resolution is only possible for thin specimens.

\begin{figure}
\includegraphics[width=0.75\columnwidth]{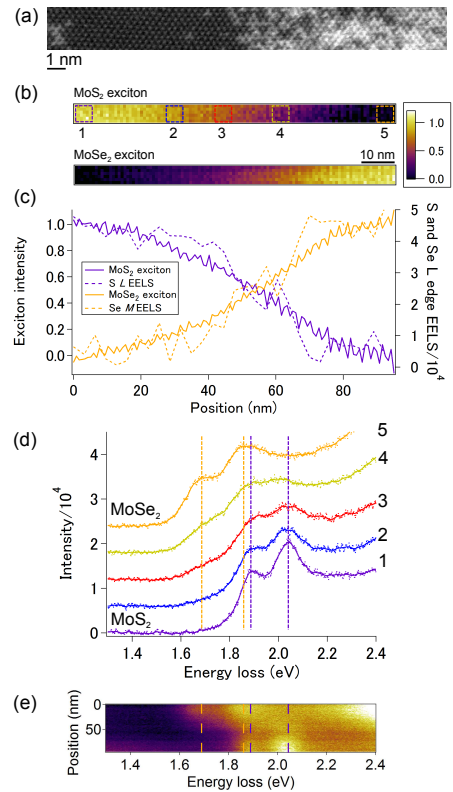}
\caption{\label{figure3} (a) Typical HAADF image of MoS$_2$/MoSe$_2$ interfaces, showing a diffuse chemical profile. (b) Maps of the fitting coefficients for MoS$_2$ (above) and MoSe$_2$ (below). (c) Comparison of the fitting coefficient profiles from with the chemical profiles measured from core-loss EELS of the S $L$ and Se $M$ edges. (d) 5 spectra integrated at different positions across the interface, with positions marked by colored squares and numbers in (b). (e) Projection of the EELS map in the spatial direction perpendicular to the interface. The change from MoS$_2$ to MoSe$_2$ is followed by peak-broadening but no apparent spectral shift.}
\end{figure}

As shown in Fig. \ref{figure3}, we have acquired 112 x 6 EEL spectra with a 0.9 nm increment across a MoS$_2$/MoSe$_2$ interface in a single-layer MoS$_{2(1-x)}$Se$_{2x}$ sample. The contribution of each material to the measured 672 spectra has been quantified using a multiple linear fit algorithm. We impose that each spectrum is a linear superposition of the two materials’ contribution. This is not necessarily true, as alloying (among other effects, such as electromagnetic coupling) may change the energy gap, as observed in photoluminescence in Mo$_{1-x}$S$_2$W$_{x}$S$_2$ \cite{Chen2013} and in MoS$_{2(1-x)}$Se$_{2x}$ \cite{Mann2014}. However, we estimated with this simple analysis the spatial extent of the region where the exciton signature changes from MoS$_2$ to MoSe$_2$. The references used have been normalized to ensure that the coefficients were between 0 and 1. In Fig. \ref{figure3} (b-c), we observe that the energy-loss spectrum changes completely from one material to the other in 50 or 60 nm. Across the interface the spectra do not show abrupt changes but suffer gradual alterations. This gradual change can be explanined by the diffuse chemical profile of the interface and also by delocalization effects. The dotted profiles in Fig. \ref{figure3} (c) show the intensities of the S $L$ and Se $M$ edges (purple and orange, respectively) measured in the same region. The intensity of these edges is proportional to the chemical composition \cite{Egerton2011}. The variation of exciton contribution in each material follows the compositional change. Atomically resolved images of typical interface is shown in Fig. \ref{figure3} (a), where a rough interface is observed, probably due to diffusion or interdiffusion during the growth process. The length scale of the region where composition changes in this typical image is smaller than in the chemical and exciton maps discussed in what follows. Yet, it exemplifies a diffuse interface at the atomic scale.

In the EELS maps we do not observe a continuous shift of the exciton energy following the chemical changes. That is, the first A exciton peak in MoS$_2$ does not continuously shift to the first A exciton peak in MoSe$_2$. In Fig. \ref{figure3} (d), 5 spectra taken from the EELS map are shown (positions marked in the map in Fig. \ref{figure3} (b)). A clear broadening of the exciton lines is observed. This indicates that in interfaces which are not wide enough when compared to the exciton wavefunction extent, the exciton energy do not follow the chemical profile. In simpler terms, even if an intermediary chemical composition is present (MoS$_1$Se$_1$, for example) in a narrow interface the exciton does not show an intermediary energy  such as observed in a homogeneous alloyed layer (1.65 eV for MoS$_1$Se$_1$\cite{Mann2014}). This behavior might be reasonably expected in luminescence experiments, where the exciton can diffuse to and recombine in the adjacent lower bandgap material. But EELS experiments measure directly the energy lost. In this case, one should be able to probe the local states within the interface.

The observation of these variations in the spectral signature in scales of the order of 10 nm indicates that the spatial resolution might indeed be better than one would expect for low loss EELS. It is known that a fast electron can lose energy creating an excitation even if `it propagates outside the material', an effect known as delocalization \cite{Egerton2011, Muller1995, Spence2006}. This effect is larger for smaller energy losses and for this reason poor spatial resolution is expected at lower energy losses. However, as remarked by J. Garcia de Abajo \cite{Abajo2010}, ``the field diverges at the origin as $~1/R$, so that large interaction contrast is expected across small distances in the region close to the trajectory''. Moreover, from a technical point of view, if the spectra of two regions are sufficiently different, spatially-resolved mapping allows discerning the two responses. An analogous behavior allowed the observation of quantum wells separated by 5 nm in cathodoluminescence experiments using spectrum-imaging \cite{Zagonel2010}.

\begin{figure}
\includegraphics[width=1\columnwidth]{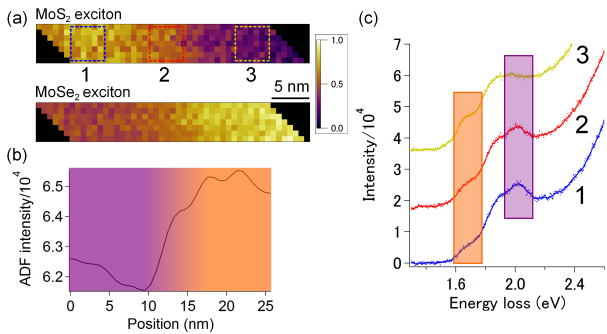}
\caption{\label{figure4} (a) Maps of the fitting coefficients for MoS$_2$ (above) and MoSe$_2$ (below) extracted using a multiple linear fit algorithm. (b) HAADF profile associated with the maps in (a) (the scales in (a) and (b) are identical). The HAADF profile, proportional to a power of the local composition, changes in a distance shorter than 5 nm along the interface. (c) 3 spectra integrated in regions across the interface. The distance between them is about 10 nm. Changes of the exciton energy-loss spectra can be observed even in this deep subwavelength regime.}
\end{figure}

In fact, in areas with smaller regions of alternating high S and Se content sharper variations of the exciton signal can be detected. In this region, the HAADF profile (Fig. \ref{figure4}(b)) shows an interface sharper than 5 nm, of the order or smaller than the size of the excitons here, in contrast to that shown in Fig. \ref{figure3}. In such regions, maps of the fitting components of the A exciton for MoS$_2$ and MoSe$_2$ in Fig. \ref{figure4} (a) show that the exciton signals follow closely the HAADF signal, which is proportional to a power of the local chemical composition($Z^\alpha$, where usually $1.4 < \alpha < 2$) \cite{Egerton2011}. Changes in spectra integrated in regions separated by about 10 nm can be seen, as exemplified by three spectra in Fig. \ref{figure4}(c), showing that deep subwavelength variations can be mapped. In these regions, the excitonic peaks are not as sharp and separable as in the larger region shown in Fig. \ref{figure3}. One possibility to interpret this result is the inherent non-local character of low energy excitations. However, this alone does not explain why the peaks are broader. A second possibility is that heterogeneous alloying leads to a broadening of the energy-loss lines, as it is known to occur in other excitonic systems \cite{Singh1985, Patane1995}. In fact, we have observed chemical heterogeneities in this sample using EELS with atomic resolution, as exemplified in Fig. \ref{figure2}. A direct correlation of local chemical heterogeneities, interface roughness and exciton behavior will surely benefit the understanding and control of these materials. 

EELS experiments have been performed on a JEOL ARM equipped with a Schottky field emission gun, a JEOL double Wien filter monochromator, and an  EEL spectrometer. A Gatan Quantum modified for low primary energy operation (15-60 keV) with higher stability was used. All EELS experiments have been performed in scanning mode using 30 keV electrons. Typical parameters of the probe were: energy profile FWHM between 20 meV and 36 meV; current 35 pA; convergence semi-angle of 12 mrad. The EELS collection semi-angle was 33 mrad. The energy dispersion of the spectrometer was set to 2 meV/channel. Samples were cooled down to about 150 K using a liquid nitrogen-cooled sample holder. EELS maps have been acquired using the dual-EELS mode, where a fast acquisition (1 ms time scale) in series with a long acquisition (100 ms to 1 s time scales) allows the observation of the unsaturated zero-loss peak and the desired spectrum. The long acquisition times used for the higher-loss spectra did not lead to a substantial energy resolution loss. The dispersion of the spectrometer was checked over long time scales showing no significant changes. Finally, the absolute energy scale has been calibrated using optical data for the MoS$_2$ exciton absorption \cite{Molina-Sanchez2013} at 1.88 eV. Atomically resolved HAADF field images have been acquired in a JEOL 2100F equipped with a cold field emission electron gun and double JEOL aberration correctors. The MoS$_{2(1-x)}$Se$_{2x}$ sample has been produced by sulfurization of MoSe$_2$ flakes at 700 \textdegree C for 3 hr \cite{Su2014}. 

To conclude we have used a monochromated electron source to spatially map the energy-loss spectra of excitons in 2D materials. We have shown that subwavelength measurements, far below the wavelength of light emitted by these excitations is possible. We have observed that spectra across diffuse MoS$_2$/MoSe$_2$  interfaces show contributions from the excitons from both materials, but no shifts. Exciton mapping at the subwavelength limit in different systems is naturally of great interest; even more if coupled to other well-establish electron microscopy techniques. This improvement in detectability of low energy excitations (due to the zero-loss tail suppression) with high spatial resolution will help elucidate different problems, such as understanding the effects of local heterogeneities in excitons confinement (in 0, 1, 2 or 3 dimensions), the origin and behavior of excitations on interfaces (such as in MoSe$_2$-WSe$_2$ \cite{Huang2014, Gong2014}),　 the abruptness of the electronic transition between metallic and semiconductor phases single layers in MoS$_2$ \cite{Lin2014, Kappera2014}, the local behavior of point defects and the influence of the local structure of dopants in their electronic properties (for example, how the luminescence of a dopant depends on the local structure of the matrix).  

\bibliography{2Dexcitons.txt}

\end{document}